\documentclass[%
reprint,
groupedaddress,
 amsmath,amssymb,
]{revtex4-2}

\usepackage{graphicx}% Include figure files
\usepackage{dcolumn}% Align table columns on decimal point
\usepackage{bm}% bold math
\usepackage{hyperref}% add hypertext capabilities
\hypersetup{
    colorlinks=true,
    linkcolor=blue,
    filecolor=magenta,      
    urlcolor=cyan,
    pdftitle={Sharelatex Example},
    bookmarks=true,
    citecolor=blue
}

\begin{document}

\preprint{APS/123-QED}

\title{A formalism for giant Goos-Hänchen shift in metasurface sensors with phase singularity}

\author{Lotfi Berguiga$^1$}%
\email{lotfi.berguiga@cnrs.fr}
\author{Sébastien Cueff$^2$}%
\author{Lydie Ferrier$^1$}%
\author{Fabien Mandorlo$^1$}%
\author{Taha Benyattou$^1$}%
\author{Xavier Letartre$^2$}%
\author{Cécile Jamois$^1$}%

\affiliation{$^1$INSA Lyon, Ecole Centrale de Lyon, CNRS, Université Claude Bernard Lyon 1, CPE Lyon, INL, UMR5270, 69621 Villeurbanne, France}%
\affiliation{$^2$Ecole Centrale de Lyon, INSA Lyon, CNRS, Université Claude Bernard Lyon 1, CPE Lyon, INL, UMR5270, 69130 Ecully, France}

\date{\today}% It is always \today, today,

\begin{abstract} 
The Goos-Hänchen (GH) shift becomes giant in resonant photonic structures, making it promising for refractive index sensors with ultimate sensitivities. %For sensors working at thecritical coupling regime, 
We provide here a complete formalism to analytically describe the GH shift and its associated sensitivity around the critical coupling regime in photonic structures. This analytical framework quantitatively connects physical parameters such as the quality factor, the angular dispersion, the beam size and the phase singularity to the GH shift. We numerically confirm this theory in two practical designs: a surface plasmon resonance sensor and a Bloch surface wave (BSW) metasurface sensor.
Coupling our theory with numerical simulations, we design a BSW metasurface whose GH sensitivity ($10^{13} \mu m/RIU$) is more than 5 orders of magnitude higher than the current state-of-the art.
We also reveal that the main practical limitation to reach 
ultimate GH sensitivities is the beam size. However, taking into account realistic beam sizes and introducing engineering dispersion for the metasurface, we calculate limits of detection for GH sensors as low as $10^{-13} RIU$ that still surpass current sensors. These results open the way for new sensing application needing high sensitivity and low limit of detection.

\end{abstract}

\maketitle

%%%%%%%%%%%%%%%%%%%%%%%%%%  body  %%%%%%%%%%%%%%%%%%%%%%%%%%
\section{Introduction}

The Goos-Hänchen (GH) effect is the lateral displacement of a reflected optical beam on an interface. This lateral displacement can reach several thousands of wavelengths in resonant photonic structures such as Fabry-Perot cavity \cite{Wang2008}, surface plasmon resonance (SPR) \cite{Yin2004,Wan2020,Zhang2021,Yan2024,Zhu2024}, 2D transition metal dichalocogenides, topological singularity \cite{You2018,GuO2020,Yan2024}, Bloch surface waves \cite{Wan2012} and Bound State in the Continuum (BIC) in metasurfaces \cite{Wu2019,Ruan2022,Du2022,Chen2023,Qi2024}. Furthermore, the amplitude of this shift is sensitive to local refractive index changes at the interface. Therefore, tailoring giant GH effect is a promising method to design very highly sensitive photonic sensors towards ultimately low limits of detection.
Indeed, exploiting giant GH shifts, sensitivities as high as $10^{6} \mu m/RIU$ and limits of detection as low as $10^{-6}-10^{-7}$ RIU have been reported experimentally \cite{Wang2008,Wan2012,Zhu2024,Li2025}, as well as detection of biomarkers with concentration as low as 0.1 fM \cite{Zhu2024}.  A key advantage of this method compared to phase sensors based on interferometry or ellipsometry  \cite{Kabashin1998,Kabashin1999,Kabashin2009,Ng2013,Li2008} lies on its much simpler instrumental development: one simply needs to measure the displacement of a light beam.\\\

Recent theoretical works predict even higher GH shifts (reaching a few mm) and GH sensitivities ($10^8 \mu m/RIU$) using various photonic resonances \cite{Zhang2021,Youssef2022,Yan2024}.
However, a complete theoretical framework for the GH shift in resonant photonic structures is still lacking. Previous works focus on specific photonic structures e.g. long range surface plasmon \cite{Chen2011}, or point out the influence of the quality factor \cite{Chen2023,Qi2024} or the critical coupling \cite{Youssef2022,Zhu2024,Yan2024} on the GH shift,
without explicitly formulating and quantifying their roles. On the experimental side, the practical limits of the GH shift remains an open question, and the optimized conditions of experimental measurements to obtain optimized GH shift and sensing are not established.

In this work, using temporal coupled mode theory (CMT), we propose a complete formalism that reveals the key parameters leading to giant GH shift and GH sensitivity for photonic structures. By approaching the critical coupling, we show how to reach ultimate GH shift and GH sensitivity with values 3 orders (resp. 5 orders) of magnitude larger than published theoretical GH shift (resp. GH sensitivity). 
Under the critical coupling conditions, we establish a direct relation between the GH shift, the lateral propagation of photons of the resonant mode and the amount of reflected light.\\
We also demonstrate the necessity to take into account the real experimental size of the light beam to design GH-based sensors. Finally, our theoretical framework provides an analytical description of the limit of detection (LOD) which is the main parameter that characterizes the efficiency of a sensor.\\% It points out the cases where approaching the  critical coupling and its phase singularity improves drastically the LOD. \\

This paper is decomposed in five parts. The first part installs the theoretical framework leading to analytical expressions for the GH shift and sensitivity in arbitrary photonic structures. The second part is the numerical validation of the theory, for two different kinds of transducers (an SPR transducer ssisted by pahse change material (PCM) and a bloch-surface wave transducer).
In the third part we show how much the beam size affects the actual GH shift. The fourth part describes and calculates the limit of detection of such sensors. The last part discusses the trade-off between ultimate sensitivity and experimental constraints to achieve it. Based on both these theoretical and practical considerations, we propose optimized, yet experimentally realistic, metasurface designs to reach ultimately low limits of detection in GH-based sensors.

\section{Formalism for Goos-Hänchen shift sensors}

\begin{figure}
    \centering
    \includegraphics[width=0.85\linewidth]{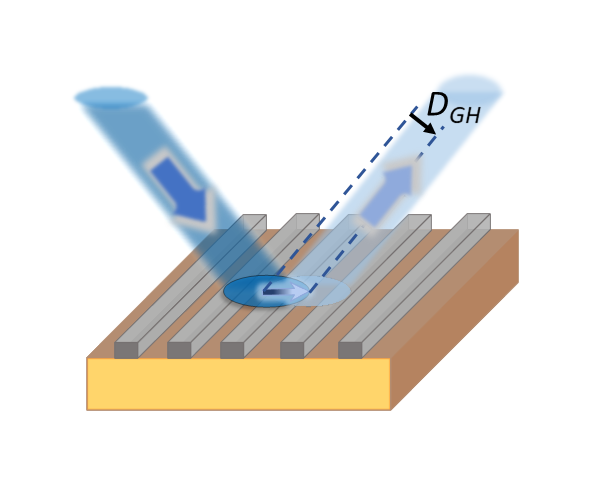}
    \caption{Goos-Hänchen effect ; lateral beam displacement of the reflected beam by a metasurface}
    \label{fig:GHshiftprinciple}
\end{figure}

When light is reflected by an interface at an angle of incidence $\theta_0$ (figure \ref{fig:GHshiftprinciple}), the lateral shift can be estimated by the expression:

\begin{equation}
D_{GH}=\frac{\int xI_r(x)dx}{\int I_r(x)dx}cos\theta_{0}
\label{eq:GHtheoriquegénérique}
\end{equation}

With $I_r(x)=E_r(x)E_r^*(x)$ the intensity spatial profile of the reflected field at plane $z=0$, corresponding to the interface.\\
This expression is valid for any kind of beam profile, and can be applied for the particular case of an infinite beam (i.e a plane wave), leading to the Artmann’s formula \cite{Polls2016}:

\begin{equation}
D_{GH}=-\frac{\lambda}{2\pi}\frac{d\phi}{d\theta}
\label{eq:GHtheorique}
\end{equation}

From eq. 2, we can infer that a phase variation of the phase $\phi$ of the reflected light would produce a large GH shift.
In the following we establish the theoretical framework for the lateral shift of an infinite beam reflected on a resonant structure, by modeling the reflectivity with CMT. 

A resonant photonic structure, such as those depicted on figure \ref{fig:GH_principe}, can be modeled by a resonator with a resonance frequency $\omega_{res}$ and an internal loss constant  $\tau_i$, coupled to a radiative loss channel, characterized by the time constant $\tau_{rad}$.  
The complex reflectivity versus the frequency $\omega$ is expressed by the following analytical relation: 

\begin{equation}
    r(\omega)=r_b\cdot \frac{\frac{1}{\tau_{rad}}-\frac{1}{\tau_i}-j\left(\omega-\omega_{res}\right)}{\frac{1}{\tau_{rad}}+\frac{1}{\tau_i}+j\left(\omega-\omega_{res}\right)}
    \label{Equ_TMC_r_Tamm1}
\end{equation}
where $r_b$ is the complex reflection coefficient of the reflecting part (Gold layer for SPR or Bragg mirror for BSW). 
Since equation (\ref{Equ_TMC_r_Tamm1}) is complex, we can derive the reflection phase as follow:

\begin{equation}
    \phi=-atan\left(\frac{\omega-\omega_{res}}{\frac{1}{\tau_{rad}}-\frac{1}{\tau_i}}\right)-atan\left(\frac{\omega-\omega_{res}}{\frac{1}{\tau_{rad}}+\frac{1}{\tau_i}}\right) + C
    \label{eq:Phaseequation}
\end{equation}

where C depends on the sign of $\tau_{rad}^{-1} - \tau_i^{-1}$: when the resonance is under-coupled, $\tau_{rad}^{-1} - \tau_i^{-1}$ is negative and $C$=0, when the resonance is over-coupled $\tau_{rad}^{-1} - \tau_i^{-1}$ is positive and $C$=$\pi$. %At critical coupling $\tau_{rad}$ equals $\tau_{i}$ and the first term in phase expression diverges at infinity. 
When plotting the reflectivity as a function of frequency, the resonance appears as an intensity dip, and the phase jump becomes increasingly sharper when approaching the critical coupling i.e. when $\tau_{rad}$ equals $\tau_{i}$ .\\

From equation (\ref{eq:GHtheorique}) and (\ref{eq:Phaseequation}), we can express the GH shift:
\begin{equation}
D_{GH}=-\frac{\lambda_{res}}{2\pi}\left(\frac{d\phi}{d\omega_{res}}\frac{d\omega_{res}}{d\theta}\right)\\
\end{equation}
\begin{equation}
D_{GH}(\theta)=\frac{4\pi acQ}{\omega_{res}^2r_{min}}\frac{1}{1 + \left(\frac{4aQ}{\omega_{res}r_{min}}\right)^2\left(\theta-\theta_{res}\right)^2}
%D_{GH}(\theta)=\frac{aQ\lambda_ {res}^2}{\pi cr_{min}}\frac{1}{1 + \left(\frac{2aQ\lambda_{res}}{\pi cr_{min}}\right)^2\left(\theta-\theta_{res}\right)^2}
\label{eq:GHshift}
\end{equation}

%%\frac{4\pi Q\lambda_{res}}{cr_{min}}
%%%

with $c$ the speed of light, $a$ the angular dispersion near  the resonance, defined as $a=\frac{d\omega}{d\theta}$, $Q$ the quality factor and $r_{min}$ the minimum of reflectivity, defined as:
\begin{equation}
    Q=\frac{\omega_{res}}{2}\cdot\frac{1}{\frac{1}{\tau_{rad}}+\frac{1}{\tau_i}} \text{\hspace{1ex}and\hspace{1ex}} r_{min}=\frac{\frac{1}{\tau_{rad}}-\frac{1}{\tau_i}}{\frac{1}{\tau_{rad}}+\frac{1}{\tau_i}}
\label{eq:Qualityfactor}
\end{equation}

\begin{figure}
    \centering
    \includegraphics[width=1\linewidth]{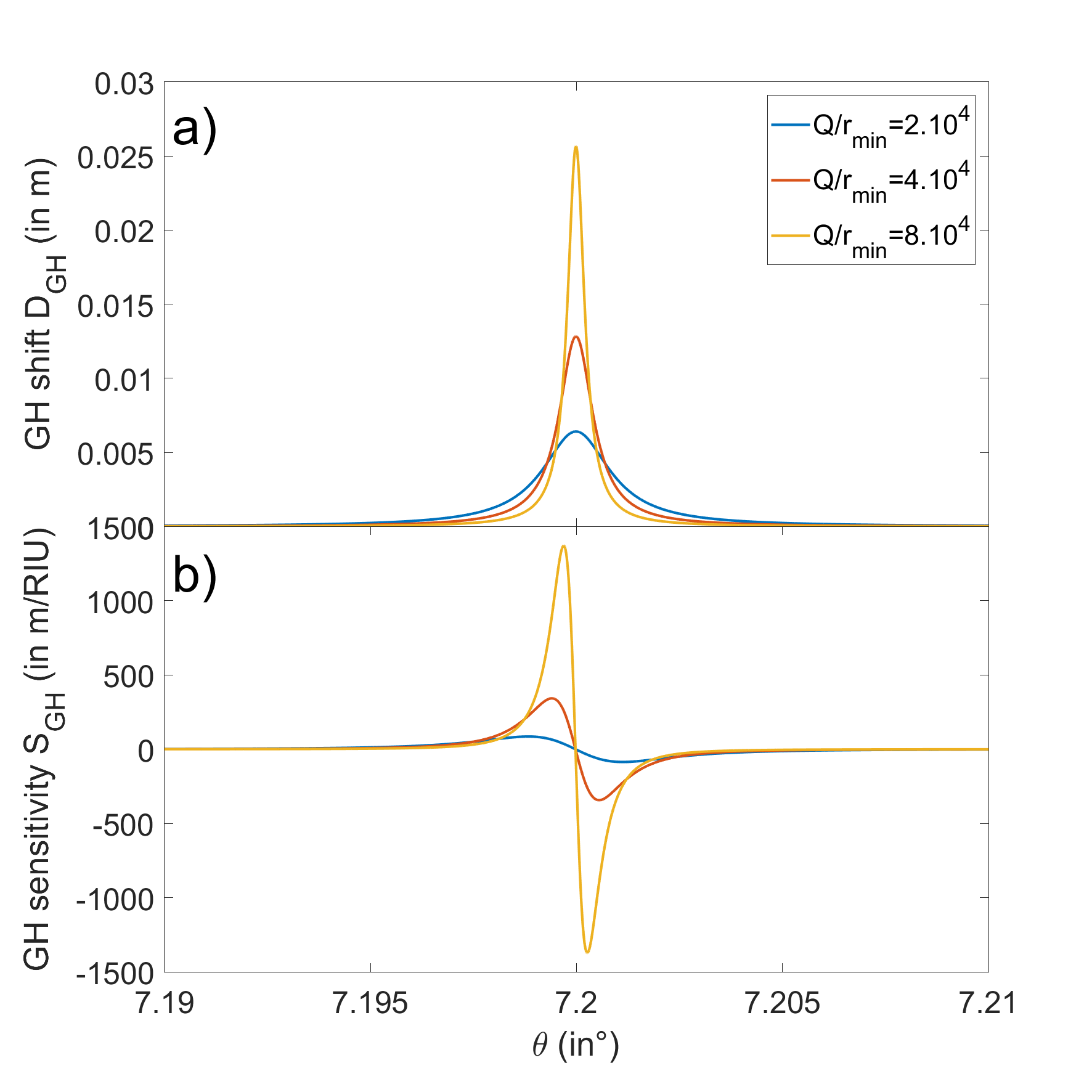}
    \caption{GH shift and sensitivity versus the incident angle for several values of $Q/r_{min}$ drawn with the temporal coupled mode theory with the following parameters, $ Q=20000$, $\lambda_{res}=1539.53 nm$, $\theta_{res}=7.20^{\circ}$ and $r_{min}=0.01~or~0.05~or~0.0025$}
    \label{fig:GHshiftandSensitivtyvsAngle}
\end{figure}

Equation (\ref{eq:GHshift}) describes the GH shift $D_{GH}$ as a Lorentzian profile with a maximal value at the resonance angle $\theta_{res}$ (figure(\ref{fig:GHshiftandSensitivtyvsAngle}a)).

If we now let the refractive index $n$ of the medium of analyte vary, we get the  bulk GH sensitivity  as a function of the angle of excitation by deriving the GH shift with respect to $n$:
\begin{align}
S_{GH}&=\frac{dD_{GH}}{dn}\label{eq:defSensibilite}\\
S_{GH}&=-\frac{\lambda}{2\pi}\frac{d}{d\omega_{res}}\left(\frac{d\phi}{d\omega_{res}}\frac{d\omega_{res}}{d\theta}\right)\frac{d\omega_{res}}{d\lambda}\frac{d\lambda}{dn}\\
S_{GH}(\theta)&=-16\frac{a^2Q^3\lambda_{res}^2}{\pi^3 c^2r_{min}^3}\frac{\theta-\theta_{res}}{\left(1+\left(\frac{2aQ\lambda_{res}}{\pi cr_{min}}\right)^2(\theta-\theta_{res})^2\right)^2}\cdot S_{\lambda}
\label{eq:SGH_vstheta}
\end{align}

With $S_{\lambda}=\frac{d\lambda}{dn}$, the sensitivity in wavelength of the transducer.\\
The GH sensitivity as a function of the incident angle, plotted from the equation (\ref{eq:SGH_vstheta}) is shown in figure (\ref{fig:GHshiftandSensitivtyvsAngle}b).
For a given wavelength, the highest sensitivity does not occur at the resonance angle $\theta_{res}$ but rather at angles $\theta_\pm$ below and above resonance, where the slope of the GH shift versus the incident angle is the most abrupt. By deriving relation (\ref{eq:GHshift}), the angle $\theta_\pm$ follows the relation: 
\begin{equation}
\theta_{\pm}=\theta_{res} \pm \frac{\omega_{res}}{2\sqrt{3}a}\frac{r_{min}}{Q}=\theta_{res} \pm \frac{\pi c}{\sqrt{3}a\lambda}\frac{r_{min}}{Q}
\label{eq:GH_angleposition}
\end{equation}

At these angles $\theta_{\pm}$, The GH shift can be written:

\begin{equation}
D_{GH}=-\frac{3a\lambda_{res}^2}{8\pi^2c}\frac{Q}{r_{min}}
\label{eq:DH_parTMC}
\end{equation}

The maximum bulk sensitivity $S_{GH}$ occurs at $\theta_{\pm}$ and is described by the formula: 
\begin{align}
S_{GH}^{max}&=-\frac{3\sqrt{3}a\lambda_{res}}{8\pi^2c}\frac{Q^2}{r_{min}^2}S_{\lambda}
\label{eq:Sensibilite_parTMC}
\end{align}

The analytical expressions in equation (\ref{eq:DH_parTMC}) and (\ref{eq:Sensibilite_parTMC}) show the  GH shift is proportional to  the ratio $\frac{Q}{r_{min}}$ and the GH sensitivity proportional to the square of the same ratio $\frac{Q}{r_{min}}$. It follows that giant GH shift and sensitivity can be obtained by either increasing the quality factor of the resonance or by lowering the minimum of reflectivity $r_{min}$. This latter condition is reached by approaching the critical coupling. This ratio $\frac{Q}{r_{min}}$ can be written as a function of the radiative and internal loss:
\begin{equation}
\frac{Q}{r_{min}}=\frac{\omega_{res}}{2}\frac{1}{\frac{1}{\tau_{rad}}-\frac{1}{\tau_{i}}} 
\end{equation}
The ratio $\frac{Q}{r_{min}}$ diverges when the time constants are equal ($\tau_{rad}=\tau_{i}$) which is the condition for critical coupling. It governs the sharpness of the phase jump at resonance. Moreover the sensitivity $S_{GH}$ increases more quickly than GH shift $D_{GH}$: the variation is quadratic with the coefficient $\frac{Q}{r_{min}}$. We have to keep in mind GH sensitivity is proportional to wavelength sensitivity $S_{\lambda}$ which traduces the spatial volume of the electromagnetic field of the mode outside the photonic structure hence accessible to sensing \cite{ElBeheiry2010}. Consequently a good GH sensor is firstly a good classical sensor (sensor interrogated spectrally or in intensity). \\
Moreover, using this formalism we can provide a more physical meaning of the GH shift. It can be related to the mean free propagation length of the photons of the resonant mode $L$ during their lifetime $\tau$ (with $1/\tau=1/\tau_i+1/\tau_{rad}$). The mean speed of the photons in the surface wave is the group velocity $v_g$. $v_g$ and $L$ can be formulated as follows:
\begin{equation}
v_g=\frac{\lambda_{res}}{2\pi \cos\theta_{res}}a~~and~~L=v_g\tau=v_g.\frac{4Q}{w_{res}}
\end{equation}
Rewriting these last expressions in  equation (\ref{eq:DH_parTMC}) leads to this new expression for the GH shift:
\begin{equation}
D_{GH}=\frac{3\cos\theta_{res}}{8}\frac{L}{r_{min}}
\end{equation}
This relation shows that the GH shift for a plane wave is governed both by the 
propagation length of the surface mode $L$ and by the critical coupling through the parameter $1/r_{min}$. Tamir and Bertoni \cite{Tamir:71} pointed out in 1971 the role of $L$ for the GH shift of a real finite beam. $L$ can be defined also as the inverse of $k_x^{''}$ the imaginary part of the wavevector of the mode. Highlighted by this approach we will later discuss the case of a finite real beam. 

\section{Numerical validation of the model in photonic sensors}

In this section we test the calculated analytical expression of GH shift and GH sensitivity with numerical simulations on two different photonic sensors: the first one based on SPR and the second exploiting Bloch Surface Waves (BSW). \\

\begin{figure}[h!]
    \centering
    \includegraphics[width=1\linewidth]{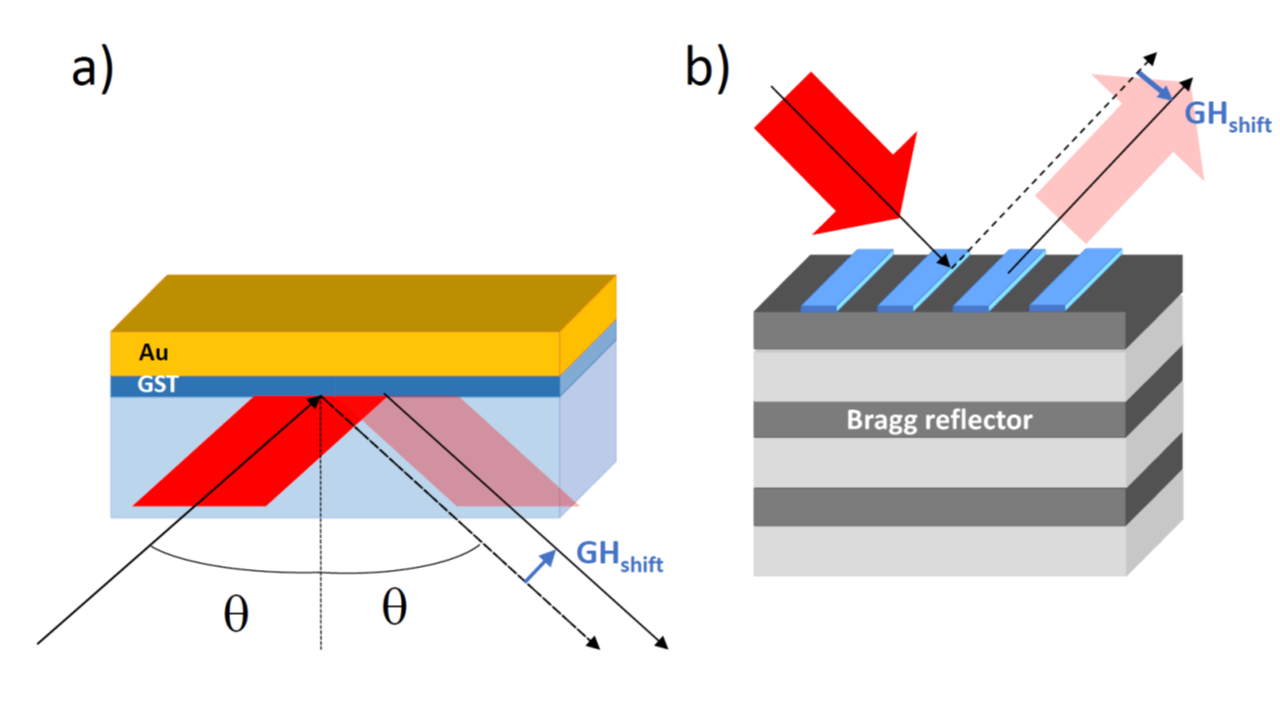}
   \caption{Goos-Hänchen effect for surface plasmon resonance (in a) and for Bloch Surface Wave metasurface (in b)). In a) between the 49.5 nm gold layer and the substrate (refractive index 1.62) a 5 nm thin film of GST is added. In b) the grating is a dielectric material of refractive index 1.8 : with width w=414 nm and period p=920 nm and height h=356 nm }
    \label{fig:GH_principe}
\end{figure}

In the following, the reflectivity versus the angle of incidence is numerically computed by  the transfer matrix method for SPR and by rigorous coupled-wave analysis (RCWA) by using a commercial software (Rsoft) for BSW metasurface. The GH shifts are calculated from the reflectivity spectra by using Artmann's equation (\ref{eq:GHtheorique}).
The GH sensitivity is computed with relation (\ref{eq:defSensibilite}) when varying the refractive index of the surrounding medium. All the physical parameters that characterize the sensor i.e. the quality factor $Q$, the sensitivity in wavelength $S_{\lambda}$ and the angular dispersion $a$ are extracted for the numerical simulation of the reflectivity when either the wavelength or the refractive index of the analyte media varies. $Q$ is calculated as the ratio between the resonant wavelength and the full-width at half maximum of the reflectivity spectra. $S_{\lambda}$ is the slope of the variation of the resonant wavelength as a function of the refractive index chnage. Finally, $a$ is the slope of the tangent in the dispersion curve i.e. the variation of $\omega_{res}$ when the angle of excitation $\theta$ varies around the angle of excitation $\theta_{res}$.\\

\begin{figure}[h!]
    \centering
    \includegraphics[width=1\linewidth]{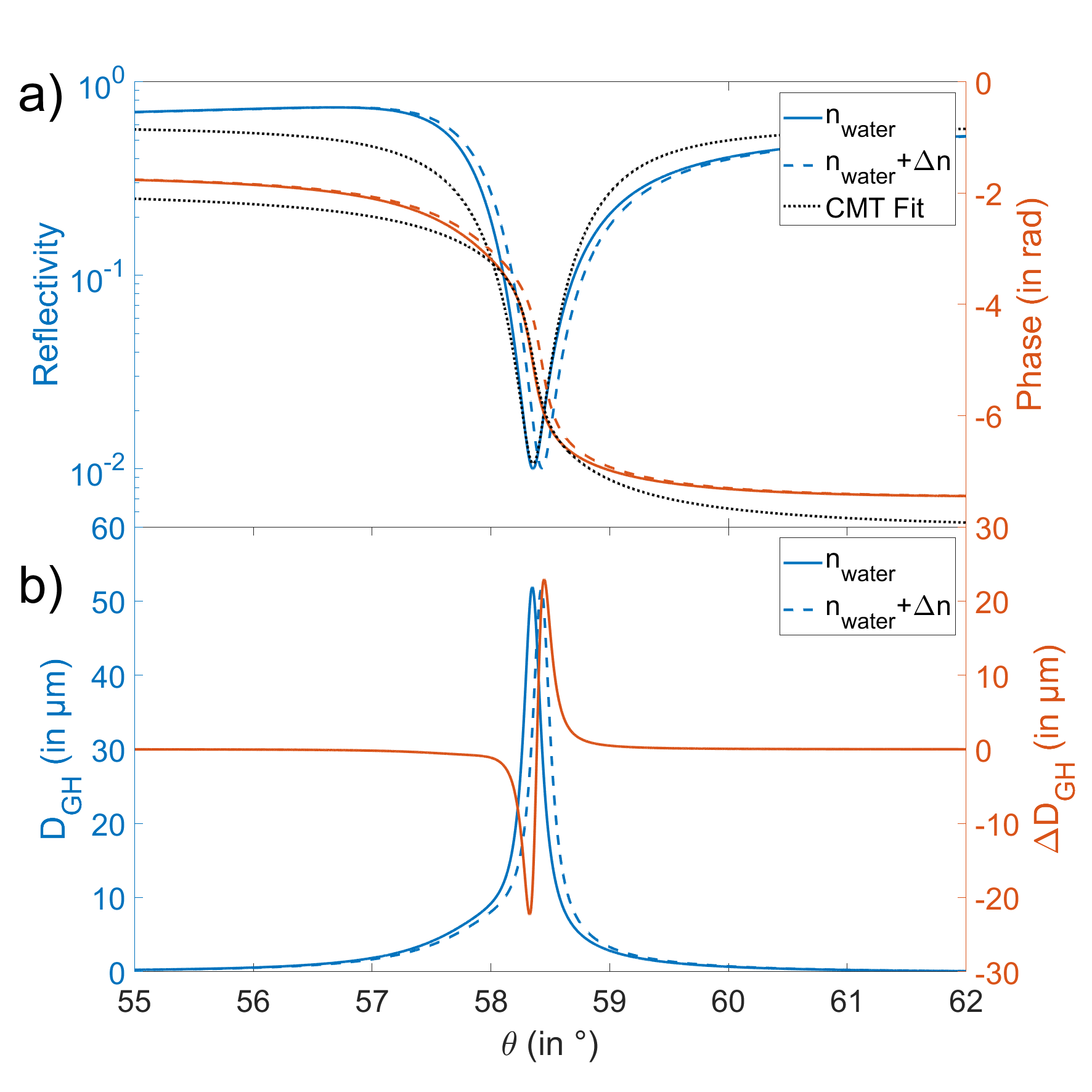}
   \caption{SPR shift due to the refractive index variation of the medium of analyte by $\Delta n = 0.001$; a) Intensity and phase of the reflectivity in the two media of analyte. The dashed black curves is the fit using the CMT.
   b) GH shift, $D_{GH}$ and variation of the GH shift $\Delta D_{GH}$ }
    \label{fig:GH_SPRPCM}
\end{figure}

The device used for the SPR sensor comprises a gold layer of 49.5 nm deposited on an amorphous GST layer of 5 nm deposited itself on dielectric glass of refractive index 1.64 (figure \ref{fig:GH_principe}a). The reflectivity versus the angle of excitation is presented at a wavelength of 750 nm. The resonance of the system appears as an intensity dip in the reflectivity spectrum at a particular angle of incidence $\theta_{res}$, as shown in figure \ref{fig:GH_SPRPCM}a). Along with this resonance, we observe an abrupt phase jump. As explained previously, the Artman's relation (\ref{eq:GHtheorique}) connects this phase variation to a GH shift in the reflected beam (see figure \ref{fig:GH_SPRPCM}b). We calculate a very large GH shift $D_{HG} > 50~\mu m$, which is a displacement that is 67 times larger than the wavelength.

To use this system as a sensor, the gold layer is put in contact with an analyte medium (water usually). When the refractive index of the analyte medium changes by a value $\Delta n$, the resonance angle is shifted, inducing a variation of the GH shift
(figure \ref{fig:GH_SPRPCM}b). For example, if we assume a variation $\Delta n$=0.001, the variation of the GH shift $\Delta D_{GH}$ is greater than $20~\mu m$, representing a value that is 26 times the wavelength and leads to sensitive sensor.\\

As displayed in figure \ref{fig:GH_SPRPCM}a), the simulated results near the resonance can be fitted in intensity and phase by the CMT model (equation (\ref{Equ_TMC_r_Tamm1}) and (\ref{eq:Phaseequation})).

\begin{figure*}[t!]
    \centering
    \includegraphics[width=1\linewidth]{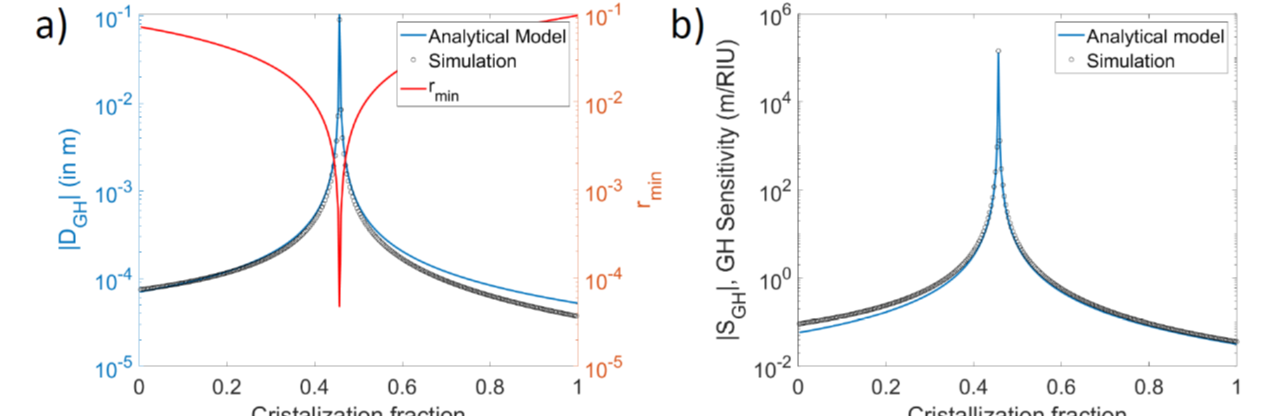}
   \caption{a) Giant Goos-Hänchen shift for SPR by crystallization of the GST layer. The blue curve is the analytical relation (\ref{eq:DH_parTMC} ) $D_{GH}=cst/r_{min}$  with $cst=\frac{3a\lambda^{2}Q}{8\pi^{}c}=5e-6$ which is determined by the physical parameters, $\lambda_{res}=750~nm$, $Q=6.5$, $a=9.62\cdot10^{15}~s^{-1}$, in black circle circle the simulation done by the transfer matrix method. In b) The GH sensitivity S$_{GH}$ with in blue line the analytical relation (\ref{eq:Sensibilite_parTMC}), with the same physical parameters $\lambda_{res}$, $Q$, $a=$, and $S_{\lambda}=4362~nm/RIU$, and in black circle the numerical simulation. GH shift and GH sensitivity are drawn in absolute value for logarithmic representation which removes the asymmetry of the phase singularity effect.}
    \label{fig:GH_Giant_SPR}
\end{figure*}

We now want to optimize the sensitivity of the sensor. As described above, our strategy is to reduce the reflectivity to a minimum, by reaching the critical coupling.  To do so we use our recently proposed concept \cite{berguiga2021ultimate}, in which we exploit the progressive change of phase of the GST layer (located between the metal and the prism, figure (\ref{fig:GH_principe})a)) to actively drive the system towards critical coupling condition via the fine-tuning of the absorption. The use of a phase change material (PCM) is more precise and easier to apply experimentally compared to metal thickness tuning that needs an angstrom precision to approach the critical coupling.\\

The figure \ref{fig:GH_Giant_SPR}a) illustrates the evolutions of the minimum of reflectivity and the associated GH shift as a function of the GST crystalline fraction. The minimum of reflectivity $r_{min}$ drops down to less than $4~10^{-5}$ for a GST crystallization fraction of $46 \%$. In this optimized system, the GH shift reaches 10 cm, which is equivalent to $133 000\lambda$. This value is 2 to 3 orders of magnitude larger than recently reported giant GH shifts \cite{Yan2024, Zhang2021}. The same figure \ref{fig:GH_Giant_SPR}a) shows that the numerically simulated GH shift $D_{GH}$ (black circles curve) are perfectly reproduced by the equation (\ref{eq:DH_parTMC}), using the physical parameters of the resonance $Q=6.5$, $a=9.62\cdot10^{15}~s^{-1}$, $\lambda=750~nm$. In figure \ref{fig:GH_Giant_SPR}b) we plot the simulated GH sensitivity as a function of the GST crystallization fraction, along with the analytical calculations from eq. (\ref{eq:Sensibilite_parTMC}) with $S_{\lambda}=4362~nm/RIU$. Here again, both numerical simulations and analytical calculations are perfectly matching. Moreover the GH sensitivity at critical coupling reaches $10^{11}~\mu m/RIU$, that is, 3 orders of magnitude larger than recently reported values in theoretical studies \cite{Yan2024}.\\

These results demonstrate the usefulness of our approach to analytically describe and optimize one specific kind of optical sensor (SPR). In order to generalize our formalism to a different platform, we extend our analysis to Bloch Surface Wave resonances.

\begin{figure}[h!]
    \centering
    \includegraphics[width=1\linewidth]{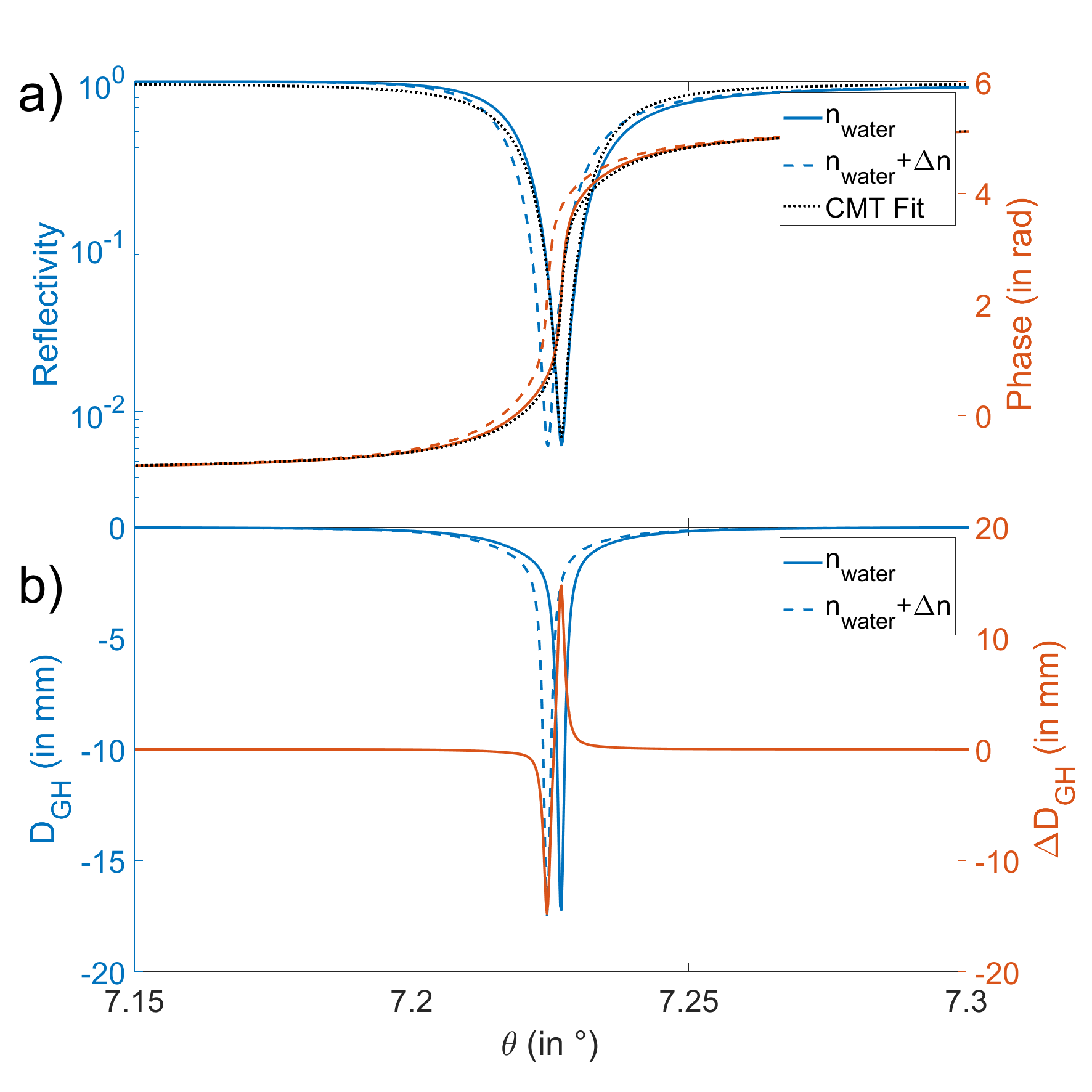}
   \caption{Bloch Surface Wave resonance shift  due the refractive index variation of the medium of analyte by $\Delta n = 6.10^{-5}$ a) Intensity end phase of the reflectivity in the two media of analyte. The dashed black curves are the fit using the CMT.
   b) GH shift, $D_{GH}$ and variation of the GH shift $\Delta D_{GH}$ }
    \label{fig:GH_BSW}
\end{figure}

In this case the transducer is a periodic dielectric grating deposited on a Bragg Mirror comprising 3 pairs of high and low dielectric layers ($Si$ and $SiO_2$). We have designed the geometry of the gratings such that the critical coupling is reached with a resonance around 1400 nm, producing a reflectivity minimum at an angle of excitation around 7° when the device is surrounded by water. The initial geometrical parameters of the grating rods are as follows: periodicity $p=~920~nm$, width $w=~414~nm$, thickness $h=~356~nm$ and refractive index 1.8.\
As shown in figure (\ref{fig:GH_BSW}), for such a design, the GH shift and its variation for a refractive index variation $\Delta n=6.10^{-5}$ are giant since their values are in the order of tens of mm i.e. 10 700 and 7100 times the wavelength, respectively. 
We compare the numerical simulations to the analytical models. As displayed with the dashed black lines in figure \ref{fig:GH_BSW} a), both the simulated intensity and phase reflectivity near resonance are well reproduced by the CMT model (equation (\ref{Equ_TMC_r_Tamm1})). \\

We now want to further approach the perfect critical coupling condition. To do so, we have added a pair of dielectric layers and modified the dimensions of the dielectric grating as follows: the period is 920 nm, the width and the thickness of the rods are 374 nm and 359 nm respectively and the refractive index of the grating is kept at 1.8.

\begin{figure}[h!]
    \centering
    \includegraphics[width=1\linewidth]{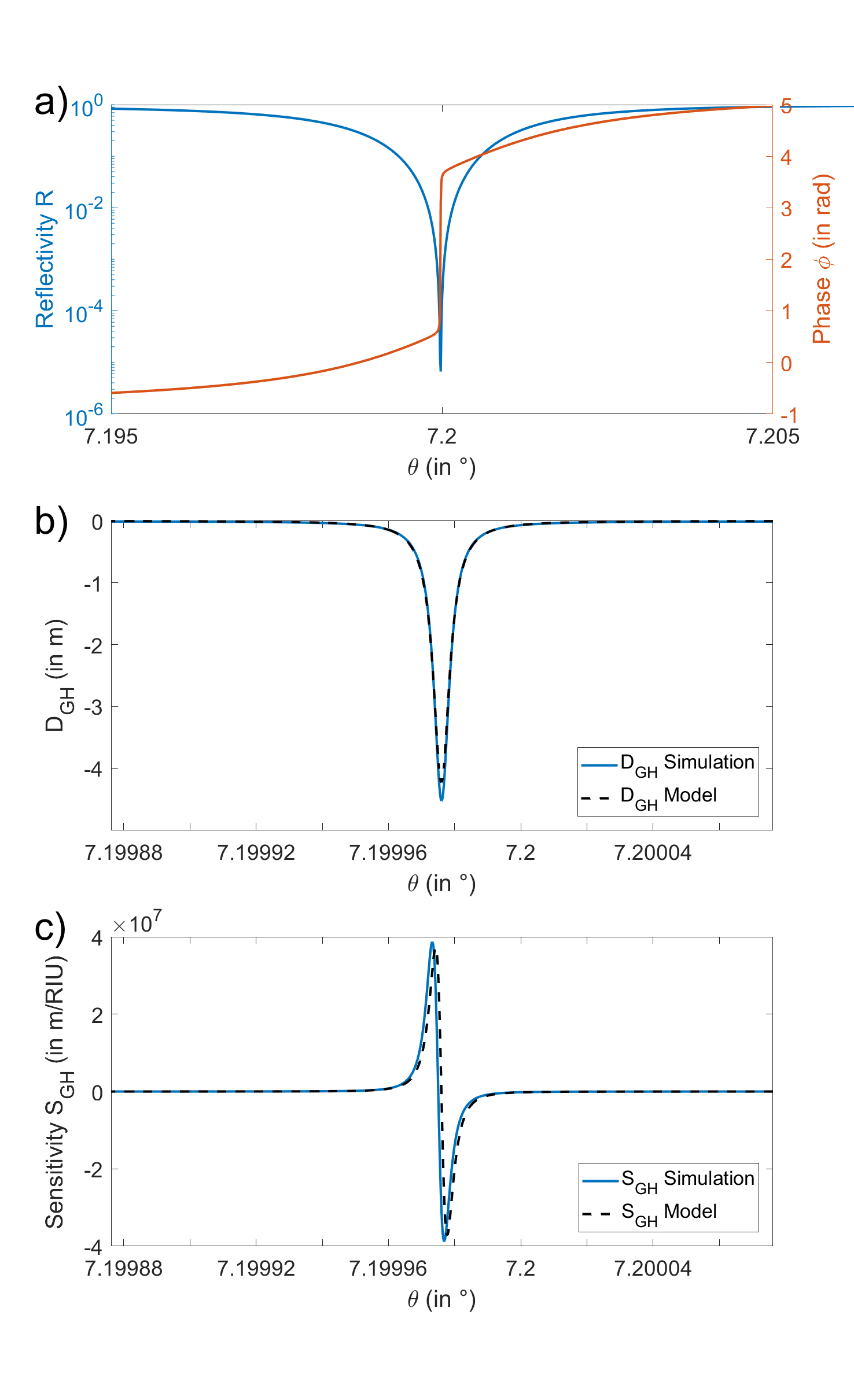}
   \caption{a) Giant Goos-Hänchen shift for BSW. In a) the numerical simulation of reflectivity and phase in TE mode for a BSW structure with a Bragg mirror of 4 pairs of layer of $Si/SiO_2$, and a dielectric grating with periodicity 920 nm. The rods width is 377.5 nm and the thickness is 359 nm. The refractive index of the grating is 1.8. In b) and c) simulation and analytical model of GH shift $D_{GH}$ and GH sensitivity $S_{GH}$ respectively when varying the refractive index of analyte medium from water to solution with $\Delta n=5\cdot10^{-8}~RIU$ }
    \label{fig:GH_Giant_BSW}
\end{figure}

As shown in figure (\ref{fig:GH_Giant_BSW}) a GH shift as high as 6 m is achieved and a GH sensitivity of $S_{GH}=3.9\cdot10^{13}~\mu m/RIU$, which is 300 000 times higher than the sensitivities reported in the literature \cite{Yan2024,Youssef2022,Zhang2021}. From figure \ref{fig:GH_Giant_BSW} a), we extract the physical parameters of the resonance: $\lambda_{res}=1399.52~nm$, the quality factor $Q=19992$ and the minimum of reflectivity at resonance $R_{min}=r_{min}^2=2.3\cdot10^{-6}$ (Such a low value is experimentally reachable, as shown in our previous work \cite{Berguiga2025}).  We then inject these values to calculate analytically the GH shift $D_{GH}$ and GH sensitivity $S_{GH}$. Here again, numerically simulated and analytically calculated values are perfectly matching throughout the range of angles. \\

These numerical simulations on SPR and BSW metasurfaces confirm that the CMT is sufficient and well suited to describe the GH shift and the GH sensitivity. It shows that giant GH shift beyond the current limits can be achieved by approaching the critical coupling and its phase singularity. Moreover, since higher quality factors are accessible with dielectric metasurfaces, these simulations clearly show the superiority of BSW metasurfaces over the SPR sensors to reach giant GH shifts and very sensitive GH sensors. Indeed, the ratio $Q/r_{min}$ is higher for BSW metasurface even without reaching the critical coupling as close as the one needed for SPR ($R_{min}$ is $2\cdot10^{-6}$ for BSW instead of $2\cdot10^{-9}$). 

%\FloatBarrier

\section{Beam size effect}\label{sec:realbeam}
\begin{figure*}[t!]
    \centering
    \includegraphics[width=1\linewidth]{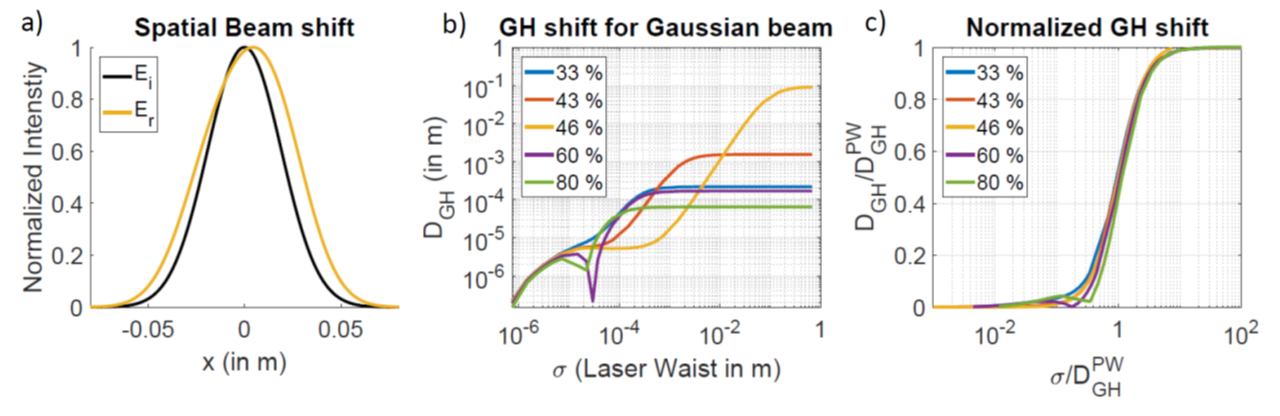}
   \caption{SPR Giant Goos-Hänchen shift for a finite Gaussian beam for different crystallization fraction of the PCM for SPR transducer. a) Gaussian beam displacement for SPR. b) Goos-Hänchen shift versus the Gaussian beam waist. c) Normalized Goos-Hänchen shift versus the normalised beam size. The factor of normalization $D_{GH}^{WP}$ is the Goss-Hänchen shift value for an infinite beam.}
    \label{fig:GH_RealBeam}
\end{figure*}

So far, our calculations and simulations assumed an infinite plane wave light source. However, realistic calculations should take into account the size and shape of the incoming beam. In the following, we provide more insights on the influence of a finite-sized Gaussian light beam on the GH effect. High GH shift requires a very precise angle of excitation with a low-divergence beam. For a Gaussian beam, the waist $\sigma$ is related to the angle divergence $\theta_0$ by the relation $\sigma=\frac{\lambda}{\pi \theta_0}$. This implies that reducing the divergence angle leads to increasing the beam size. In other words, the GH shift definition provided by equation (\ref{eq:GHtheorique}) is only true for beams with infinite diameters (the limit of a plane-wave). Wan and Zubairy \cite{Wan2020} studied numerically the effect of the size of a Gaussian beam on the GH shift for SPR sensors and showed a reduction of the GH shift when the beam size decreases. Tamir and Bertoni \cite{Tamir:71}, showed theoretically and numerically that for a Gaussian beam, the lateral shift reaches a limit that is related to the inverse of the imaginary part of the wavevector $k_x^{''}$ of the resonant mode, in other word the lateral propagation distance $L=1/|k_x^{''}|$ of the surface wave. They showed the GH shift is truncated if the beam size is smaller than $L$. They also showed when the beam size is far greater than $L$, the GH shift reaches a limit.
In figure (\ref{fig:GH_RealBeam}) we show similar results. 

To compute the GH shift for a realistic beam with equation (\ref{eq:GHtheoriquegénérique}), the 1D spatial incident electric field $E_{i}(x)$ of a Gaussian beam with a waist $\sigma$ and an angle of incidence $\theta_{0}$ at plane $z=0$   can be decomposed in angular spectrum (i.e. in plane waves)\cite{Wan2020}:

\begin{equation}
E_i(x)=\frac{1}{\sqrt{2\pi}}\int_{-\infty}^{\infty}\tilde{E}(k_{x})e^{ik_xx}dk_x
\end{equation}

where 

\begin{equation}
\tilde{E}(k_{x})=\frac{\sigma_x}{\sqrt{2}}exp\left(-\frac{\sigma_x^2(k_x-k_{x0})^2}{4} \right) 
\end{equation}

with $\sigma_{x}=\frac{\sigma}{cos\theta_{0}}$ the projected waist  on the interface, $k_x=\frac{2\pi n_0}{\lambda}sin\theta$ and $k_{x0}=\frac{2\pi n_0}{\lambda}sin\theta_0$.

The reflected beam en-goes the reflection $r(\theta)=r(k_x)$ and can then be written as:
\begin{equation}
E_r(x)=\frac{1}{\sqrt{2\pi}}\int_{-\infty}^{\infty}r(k_x)\tilde{E}(k_{x})e^{ik_xx}dk_x
\end{equation}

In the following, we calculate the effect of the beam size on the GH shift in the previously optimized SPR sensor. As shown in the figure(\ref{fig:GH_RealBeam}b), the GH shift increases when the critical coupling is approached for a fraction of crystallization around $46 \%$. From figure (\ref{fig:GH_RealBeam}b), for each phase state of the PCM, the GH shift increases with the size of the beam and always reaches a plateau. Each plateau corresponds to the theoretical value found with the relation (\ref{eq:GHtheorique}) i.e. for an infinite beam. It confirms that the GH effect is attenuated if the real beam size is small. A generalized trend can be found when both the GH shift of the real beam and the waist of the laser are normalized by the GH shift of the infinite beam (i.e. obtained with Artmann's equation (\ref{eq:GHtheorique})) and noted $D_{GH}^{WP}$. The normalized GH shift value versus the normalized waist of the laser is displayed in figure (\ref{fig:GH_RealBeam}c). With this normalization, all the curves for the different coupling conditions -- hence for the different GH shift -- are mostly superposed. More strikingly, the normalized value is limited to one, meaning that even for giant GH shifts the total displacement only represents a fraction of the beam size. Or in other words, the displacement cannot be bigger than the size of the beam. This fact, although obvious, is usually omitted in GH effect experiments and the size of the beam is not systematically mentioned despite its basic role on the GH effect. Tamir and Bertoni came to similar conclusions \cite{Tamir:71} and we generalized it for photonic structures working at critical coupling. 

From these normalized curves, it can be concluded that if we do not want to attenuate the GH shift by more than 40\%, the real beam should have a waist at least equal to the GH shift of the infinite beam i.e. the diameter should be twice bigger than the maximal GH shift. And if we want to limit the attenuation to only $20\%$, the beam diameter should be at least 4 times greater than the maximal GH shift. For example, for the GST crystallization fraction of $43 \%$ where the critical coupling is the nearest, the GH shift is 1 mm for an infinite beam. Consequently, the real beam should be at least 4 mm in diameter to get a real GH shift of 0.8 mm. And for crystallization fraction of $46 \%$ the width of beam should be 36 cm to obtain a GH shift of 7.2 cm\\% Indeed we can see in In figure(\ref{fig:GH_RealBeam}c) that GH shift for BSW metasurface follows the same normalized curve as SPR.

From this calculations, we clearly show that it is essential to take into account the GH shift attenuation due to the finite beam size in experimental measurements. This limits the sensor efficiency and its field of application, since a centimeter-scale beam is unrealistic for many applications. This is the main limitation to reach the highest possible GH shift and sensitivity. However, in the last part of the manuscript we will discuss how to mitigate this constraint by optimization and reach very high sensitivity.\\

\section{Limit of detection for sensing}
For sensing applications, the limit of detection (LOD) characterizes the efficiency of the sensor. This LOD is governed both by the limit of resolution of the setup and by the sensitivity of the transducer (the photonic structure). It is the smallest GH shift variation detectable from noise with a signal noise ratio of 3. More precisely, for transducers based on the GH effect, the LOD is the ratio between the smallest measurable displacement $\Delta x$ (with a signal noise ratio of three) and the GH sensitivity $S_{GH}$ :
\begin{equation}
    LOD_{GH}=3\frac{\Delta x}{S_{GH}}
    \label{eq:LOD}
\end{equation}

The smallest measurable displacement $\Delta x$ is directly connected to the optical noise of the detector \cite{Barnett2003}. Here, we consider detecting the beam displacement with a two quadrant photodiode, measuring the difference of intensity arriving at each part of the quadrant. The shot noise, which is the quantum limit, provides the minimum detectable displacement $\Delta x^{ql}$ related by the number of photon $N$ received by the detector and the beam waist $\sigma$ \cite{Barnett2003}:

\begin{equation}
    \Delta x^{ql}=\frac{\pi^{1/2}\sigma}{2}\frac{1}{N^{1/2}}
\label{eq:Deltax}
\end{equation}

 For a wavelength $\lambda$, and an integration time $\tau$, the number of detected photons reflected by the GH photonic sensor depends on the laser power $P_{laser}$  and the reflectivity $R(\theta_{\pm})=4R_{min}=4r_{min}^2$:

\begin{equation}
   N=\frac{4\lambda_{res} \tau}{hc}\cdot R_{min}P_{laser}
   \label{eq:NbrPhoton}
\end{equation}
With $h$ the Planck's constant. We have to keep in mind that the waist of the laser $\sigma$ must be at least 2 times greater than the GH shift. If we assume that $\sigma=2D_{GH}$, 
by replacing $D_{GH}$ from equation (\ref{eq:DH_parTMC}) and N from equation (\ref{eq:NbrPhoton}) into equation (\ref{eq:Deltax}), $\Delta x^{ql}$ becomes:
\begin{align}
    \Delta x^{ql} &=\frac{3}{16}\left(\frac{h\lambda_{res}^3}{\pi^2 c\tau P_{laser}}\right)^{\frac{1}{2}}\frac{aQ}{r_{min}^2}\\
    \Delta x^{ql} & \propto\frac{Q}{r_{min}^2}
\label{eq:Deltax_TMC}
\end{align}

It turns out that when the critical coupling is approached i.e. when the minimum reflectivity $r_{min}$ decreases, the smallest measurable displacement $\Delta x^{ql}$ increases.

Injecting $\Delta x^{ql}$ and $S_{GH}$ from equation (\ref{eq:Sensibilite_parTMC}) into equation (\ref{eq:LOD}), the LOD at the quantum limit can be written as:
\begin{align}
    LOD_{GH}^{ql} &=\left(\frac{3\pi}{4N_{laser}}\right)^{\frac{1}{2}}\frac{\lambda_{res}}{S_{\lambda}Q}\\
    LOD_{GH}^{ql} &=\left(\frac{3\pi}{4}\frac{hc}{\lambda_{res}}\frac{1}{\tau P_{laser}}\right)^{\frac{1}{2}}\frac{\lambda_{res}}{S_{\lambda}Q} \label{eq:LOD_TMC_long} \\ 
    LOD_{GH}^{ql} & \propto\frac{1}{S_{\lambda}Q}
\label{eq:LOD_TMC}
\end{align}

with $N_{laser}$ the number of photon emitted by the laser during integration time $\tau$.\\
 It leads to the fact that the LOD at quantum limit does not depend on the critical coupling and its phase singularity. In other words the critical coupling has no effect on the LOD at quantum limit: it does not improve it. 
However, such small displacement values $\Delta x^{ql}$ are in practice very hard to reach (they are in the $pm$ to $\mathring{A}$ range) and could easily be perturbed by other experimental sources of noise.  For an experimental measurement with the resolution of displacement $\Delta_x\gg\Delta x^{ql}$, the LOD is equal to :
\begin{equation}
LOD_{GH}=\frac{8\pi^2c}{\sqrt3}\frac{1}{a\lambda_{res} S_{\lambda}}\left(\frac{r_{min}}{Q}\right)^2\Delta x
\label{eq:LOD_badlimit}
\end{equation}

In this case, it would be very interesting to approach the critical coupling, i.e. decrease $r_{min}$, as $LOD_{GH}$ is proportional to $r_{min}^2$. It is then possible to reach very low detection limit despite high experimental noise thanks to the critical coupling approach.

\section{Discussion}

\begin{table*}[t!]
  \centering
\begin{tabular}{cccccc}
\\
\hline
$D_{laser}$  & $R_{min}$ & $S_{GH}$ ($\mu m/RIU$) & $\Delta x^{ql}$ (nm)  & $LOD_{GH}$ (RIU)& $LOD_{GH}^{ql}$ (RIU) \\
\hline
$200~\mu m$  & 0.007 & $1.6\cdot10^7$ & $1.32\cdot10^{-3}$ & $2.32\cdot10^{-8}$ & $2.6\cdot10^{-13}$ \\
$2~mm$ & $6.9\cdot10^{-5}$ & $1.5\cdot10^9$ & $1.32\cdot10^{-1}$ & $2.32\cdot10^{-10}$ & $2.6\cdot10^{-13}$ \\
$1~cm$ & $2.7\cdot10^{-6}$ & $3.8\cdot10^{11}$ & $3.3$& $9.32\cdot10^{-12}$ & $2.6\cdot10^{-13}$\\
$5~cm$ & $1\cdot10^{-7}$ & $1\cdot10^{13}$ & $82.7 $& $3.7\cdot10^{-13}$ & $2.6\cdot10^{-13}$\\
\hline
\end{tabular}
\caption{GH sensitivity and LOD for an metasurface that would work at 450 nm with a flat-band structure (angular dispersion $a=1.6\cdot10^{13}~s^{-1})$ and a quality factor $Q=30 000$. Two LOD are computed. The first one is LOD for an experimental limit of measurable displacement $\Delta x_{exp}$ equal to $120~nm$ as reported in reference \cite{Zhu2024}. The second one, $LOD_{GH}^{ql}$  is related to the smallest measurable displacement at the quantum noise limit $\Delta x^{ql}$ for a laser power of 20 mW and an integration time of 1 s. The laser beam diameter is $D_{laser}$ and is the starting point for all the other parameters of the table. We fix $D_{laser}=4D_{GH}$.}
  \label{tab:one}
\end{table*}

\begin{table*}[t!]
  \centering
\begin{tabular}{cccccc}
\\
\hline
kind of Work  & references & $D_{GH}$ & $S_{GH}$ ($\mu m$/RIU)  & LOD (RIU) & Mechanism \\
\hline
Experimental  & Wan et al. \cite{Wan2012} & 750 $\mu m$ & $3.4~10^{6}$ & N.E. & BSW \\
Experimental & Zhu et al. \cite{Zhu2024} & 200  $\mu m$ & $1.7~10^{5}$ & $7. 10^{-7}$ & SPR assisted by PCM  \\
Theoretical& Qi et al. \cite{Qi2024} & 5000 $\lambda$  & N.E. & N.E. & BIC and merged BIC\\
Theoretical & Yan et al.\cite{Yan2024} & 15 mm & $1.3~10^{8}$ & NE & SPR assisted by 2D TDM\\
Theoretical & Youssef et al. \cite{Youssef2022} & 2 mm & $5~10^{7}$ & N.E. & SPR assited by PCM\\
Theoretical & Zhang et al. \cite{Zhang2021} & 2000 $\lambda$ & $4~10^{7}$ & N.E. & Surface Phonon Resonance and LSphR\\
Theoretical & This work & $2~10^6~\lambda$ & $10^7~to~10^{13}$ & Down to $10^{-13}$ & BSW or Phase singularity with flat band\\
\hline
\end{tabular}
\caption{Comparison of this work with the scientific literature. N. E means Not Evaluated.}
  \label{tab:two}
\end{table*}

We have shown in this work that there is no limit to the lateral displacement of a reflected beam by the GH effect for resonant photonic systems working at the critical coupling regime. In this work, for the first time to the best of our knowledge, the physical parameters that rule the GH shift as well as the GH sensitivity are provided. We have shown with equation (\ref{eq:DH_parTMC}) and (\ref{eq:Sensibilite_parTMC}) for GH shift and GH sensitivity the special role of the ratio $Q/r_{min}$. The GH sensitivity evolves much faster than the GH shift when approaching the critical coupling condition and its phase singularity (i.e. when increasing the ratio $Q/r_{min}$). It leads to sensor far more sensitive than any other method. However we also demonstrated that sensors with GH shifts beyond current limits are not useful since ultimate sensitivity requires unrealistically large beam sizes (several tens of centimeters to meters). A solution to tackle this issue is to find optimized trade-offs between high sensitivity and realistic beam size. In this quest, we make use of the two following properties: i) the GH shift varies more slowly than GH sensitivity when $Q/r_{min}$ is increased and ii) the GH shift diminishes faster than the GH sensitivity when the wavelength of resonance $\lambda$ decreases.
To better understand this behavior between GH shift and GH sensitivity, we introduce a new figure of merit $FOM_{GH}$ defined as follow:

\begin{align}
FOM_{GH}&=\frac{S_{GH}}{D_{GH}}\\ %\label{eq:FOMdef}
FOM_{GH}&=\frac{\sqrt{3}}{\lambda_{res}}\frac{Q}{r_{min}}S_{\lambda} 
\label{eq:FOMrel}
\end{align}

From the definition of $FOM_{GH}$, between two sensors with the same GH shift, the best one has a better GH sensitivity, and if we compare two sensors with the same GH sensitivity the better sensor is the one with the lower GH shift. The objective is to improve the figure of merit of a GH sensor. To achieve this, we must increase its GH sensitivity while decreasing its GH shift. As shown by the equation (\ref{eq:FOMrel}), the $FOM_{GH}$ is enhanced by reducing the resonance wavelength $\lambda_{res}$ and increasing the ratio $Q/r_{min}$. It is also worth mentioning from the same equation that a better sensor has greater wavelength sensitivity $S_{\lambda}$. Although the angular dispersion $a$ does not appear in the figure of merit relation, it is better to reduce its value. Actually when increasing the ratio $Q/r_{min}$ leads to a drastic enhancement of the GH sensitivity $S_{GH}$ but in the same time if we want to keep at least the GH shift constant in other words the product $a.Q/r_{min}$ constant (see equation (\ref{eq:DH_parTMC})), it will be necessary to decrease $a$.

The rationale we propose to design a transducer with high sensitivity is as follows: one should start from the expected beam size of the measurement setup and 
subsequently optimize the design by getting high quality factor, lowering the dispersion angle $a$ and
lower as much as possible the reflectivity by working at the critical coupling regime. This criteria leads to the conclusion that quasi-BIC metasurfaces with i) a quasi flat band  ii) in the critical regime iii) at wavelength in the blue domain are very good candidates for very highly sensitive GH sensor.
Xin Qi et al. \cite{Qi2024} proposed theoretically quasi-BIC metasurface with criteria i). They reported, metasurface with both a high quality factor and a quasi flat-band with an angular dispersion $a$ lower than $6\cdot10^{13}~s^{-1}$ which is 150 times lower than the value provided in our work for BSW and SPR transducer.\\ 

In table (\ref{tab:one}) we illustrate some GH sensitivity and LOD for a metasurface working at the critical coupling, with this value of flat band, with a resonance at 405 nm and a quality factor equal to 30 000. The expected diameter in the experiment determines all the values in the table. We see that the $LOD_{GH}^{ql}$ determined by the quantum limit reaches in every cases $2.6\cdot10^{-13}~RIU$. As explained above, this LOD (see equation \ref{eq:LOD_TMC_long}) does not depend on the critical coupling. However, as explained above, for small beam diameters it is very hard to reach such small displacement values $\Delta x^{ql}$. In table (\ref{tab:one}) these LOD values show this evolution with the minimum of reflectivity when it is assumed that the minimum detectable displacement is 120 nm as reported in a recent experimental work \cite{Zhu2024}. It turns out very low LOD can be reached by approaching the critical coupling without great effort on the experimental $\Delta
 x$ value. \\
Moreover, even for small beam the LOD is very low compared to other methods of sensing. In this case, the critical coupling acts as an amplifier of the sensitivity, hence improving the LOD.
In any case, whether we are limited or not by the quantum limit, such low LOD opens the way to new sensing applications such as gas sensing, quantum sensing and metrology \cite{Shui2019,Khan2020,Asiri2023}. Actually the table (\ref{tab:two}) compares this work to the literature and show the improvements it can bring in the sensing domain.

\section{Conclusion}

With the temporal coupled mode theory, a formalism for the Goos-Hänchen effect of resonators working at the critical coupling regime is provided and has been verified numerically on canonical examples, namely SPR and BSW based transducers. This formalism allows to reveal that there is no limit in the Goos-Hänchen shift and it will be possible to design beyond giant Goos-Hänchen shift by approaching the critical coupling and its phase singularity. By using the same formalism, the sensitivity of Goos-Hänchen sensors is described analytically and leads to ultimate sensitivity at the critical coupling (equations (\ref{eq:DH_parTMC}) and (\ref{eq:Sensibilite_parTMC})). It lifts the lid about the possibility of Goos-Hânchen effect for ultimate sensitivity when metasurface are considered instead of the usual surface plasmon resonance. In this work, it is also pointed out the basic role of the real beam size on the Goos-Hänchen shift. This is the main constraint and limitation of Goos-Hänchen sensing applications. To mitigate this, we propose a trade-off to achieve high sensitivity with realistic beams by introducing a figure of merit $FOM_{GH}$. Actually, by determining all the physical parameters involved in the Goos-Hänchen effect and in the figure of merit, this works provides a roadmap to design optimal metasurfaces as the most sensitive sensor reaching ultimate detection limit. This optimal sensor should be a quasi-BIC metasurface with ultra flat-band working at the critical coupling. This works also proves theoretically that when the limit of detection is governed by the quantum limit, the critical coupling plays no role (equation (\ref{eq:LOD_TMC_long})). However in case of more realistic experimental cases the critical coupling approach plays as an amplifier (equation (\ref{eq:LOD_badlimit})) for sensitivity hence an attenuator for the limit of detection. Limit of detection as low as $10^{-13}$ RIU can be reached. Such low limit opens the way for new sensing applications, in  biodetection,  gas sensing,  quantum sensing and metrology.

%%%%%%%%%%%%%%%%%%%%%%% References %%%%%%%%%%%%%%%%%%%%%%%%%

%%%%%%%%%% If using BibTeX:
\bibliography{biblio_GHshift}

\end{document}